\documentclass[final,1p,times]{elsarticle}
\usepackage{amsmath,amssymb,hyperref}
\journal{International Journal of Non-Linear Mechanics}

\def\p{\partial}
\def\bx{\bgroup \bf x\egroup}
\def\br{\bgroup \bf r\egroup}
\def\bn{\textbf{n}}
\def\bi{\textbf{e}}
\def\ve{\varepsilon}

\def\be{\begin{equation}}
\def\ee{\end{equation}}
\def\const{\mathop{\rm const}\nolimits}

\let\Im\I

\begin{document}

\begin{frontmatter}



\title{Thick smectic shells}


\author{O. V. Manyuhina and M. J. Bowick} 
\address{Physics Department, Syracuse University, Syracuse, NY 13244, USA}
\begin{abstract}
The known ground state of ultrathin smectic films confined to the surface of a sphere is described by four +1/2 defects assembled on a great circle and a director which follows geodesic lines. Using a simple perturbative approach we show that  for thick smectic films on a sphere with planar anchoring this solution breaks down, distorting the smectic layers. The instability happens  when the bend elastic constant exceeds the anchoring strength times the radius of the inner sphere.  Above this threshold, the formation of a periodic chevron-like structure, observed experimentally as well, relieves geometric frustration. We quantify the effect of thickness and curvature of smectic shells and  provide insight into the wavelength of the observed texture. 
\end{abstract}

\begin{keyword}
geometric frustration \sep smectic liquid crystals \sep periodic structure; 
\end{keyword}

\end{frontmatter}

\section{Introduction}

\begin{figure}[tb]
\centering
\includegraphics[height=50mm]{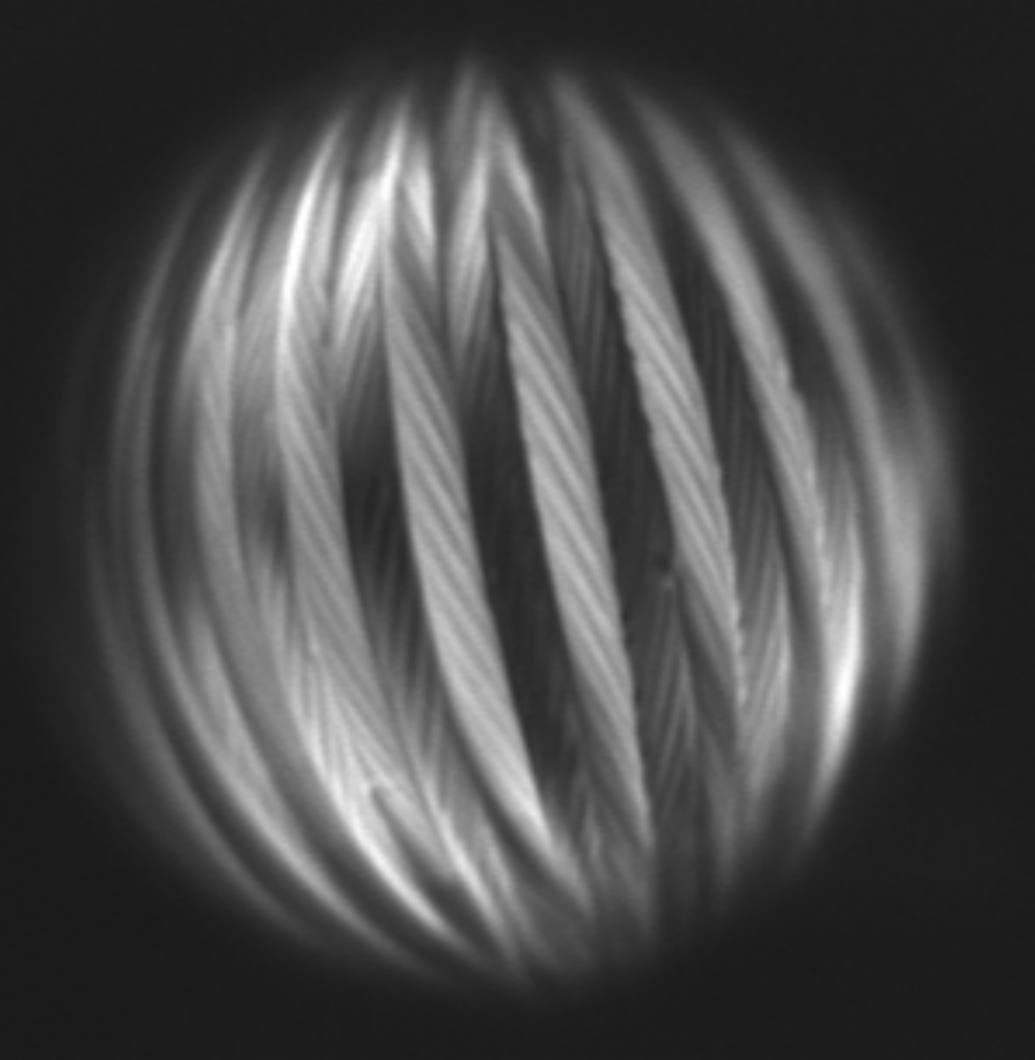}
\caption{Birefringence texture observed in experimental smectic shells~\cite{nieves:2012,nieves:2011} with radius $R=92~\mu$m and thickness $h=3.4~\mu$m.   Image courtesy of Teresa Lopez-Leon and  Alberto Fernandez-Nieves.}
\label{fig:alberto}
\end{figure}

Thin liquid crystal films, coating spherical surfaces, exhibit chevron-like periodic instability patterns, as in Fig.~\ref{fig:alberto}~\cite{nieves:2012,nieves:2011,rudquist:2011}, when observed under an optical polarizing microscope. The mechanism of this pattern formation is two-fold. On the one hand, cooling the nematic 8CB (4-$n$-octyl-4-cyanobiphenyl)  phase  towards the smectic phase with the  transition temperature $T_{\rm N-SmA}=33.5^\circ$C results in formation of smectic layers, orthogonal to the spherical surface, which locally tend to maintain constant spacing.  On the other hand,  because of the  underlying curvature of a sphere and finite thickness of the liquid crystal film, this constraint cannot be satisfied globally, leading  to geometric frustration~\cite{book:sadoc,kamien:2009}.  In  smectic-like systems geometric frustration  can be relieved by introducing topological defects, such as  focal conics, dislocations or curvature wall defects~\cite{book:intro,blanc:1999} or forming spatially periodic texture, for example, during embryogenesis in presence of  intestinal tissue growth~\cite{martine:growth}. However, a quantitative description of energy minimizing space-filling  smectic structures still remains challenging, because of the intricate connection between curvature of the embedded space, local energetic constraint of equally spaced layers,  global topological constraints as well as the  boundary conditions. In this paper we propose a simple but realistic approach to study the  effects of thickness, curvature and elastic properties of smectic shells on the 
characteristic wavelength of  instability texture. 
%%%growth of the bend elastic constant $K_3$ and the expulsion of bend distortions from the sample~\cite{cladis:1974}, similar to the Meissner effect in superconductors.

 %%\mpar{please check caption Fig 1}

Nematic shells with in-plane two-dimensional (2D) order have been extensively studied before both experimentally and theoretically~\cite{vitelli:2006,bowick:review,napoli:2012}. Smectic shells can be treated as the limiting case of nematic shells when the bend elastic constant $K_3$ is much larger than splay elastic constant $K_1$ in the Frank free energy~\cite{frank:1958}. The anisotropy of elastic constants  $K_3\neq K_1$ influences the equilibrium arrangement of  topological defects in spherical nematics~\cite{bowick:prl}  as well as the equilibrium shape of closed vesicles with in-plane orientational order~\cite{bowick:pnas}. On a curved substrate, an infinitely thin (2D) liquid crystal film with the nematic director $\bn$ aligned along geodesics $(\bn\cdot\nabla)\bn=0$~\cite{kamien:2009,bowick:prl,book:docarmo} contains no bend distortions, and is thus the ground state in the limit  $K_1\ll K_3$. For a sphere the geodesics coincide with great circles, and thus the director $\bn$ is preserved under parallel transport along meridians, resulting in splay-rich equilibrium textures~\cite{bowick:prl}. The question we pose here is how this texture changes when we allow the smectic film on a sphere to have a finite constant thickness. The increase of dimensionality of the system from 2D to 3D results in the dilation of smectic layers at the outer spherical shell, which can be compensated by the insertion of extra layers (dislocations) or by bending of the layers (curvature walls). Experimental data~\cite{nieves:2012,nieves:2011,rudquist:2011} suggests that geometric frustration in thick smectic shells is resolved by curvature walls rather than dislocations (see Fig.~\ref{fig:alberto}). %For planar geometry the buckling of smectic layers, known as the Helfrich--Hurault instability, can be driven by the mechanical dilative strain~\cite{book:intro,napoli:2009} rather than the curvature of the substrate. 

In this paper we distinguish (i)~`ultrathin' smectic films with director field following the meridians of an inner sphere, which remain stable under extension into the third dimension along the radius and (ii)~`thick' smectic shells when the tilt of the director out of the tangent plane of the spherical surface (bending of the layers) is energetically favoured. Similar to the Barbero--Barberi criterion for thin nematic films~\cite{BB:1983}, here we find that the configuration (i) becomes unstable when $K_3 > R W_a$, where $R$ is the radius of an inner sphere and  $W_a$ is the anchoring strength of the outer spherical shell.  Increasing the thickness $h$ of the smectic film above the critical value ($\propto K_3/W_a -R$) one might expect  periodic undulations of smectic layers  resulting in the formation of crescent-like domains as in Fig.~\ref{fig:alberto} and in Refs.~\cite{nieves:2012,nieves:2011,rudquist:2011}. Thus the relationship between thickness~$h$ and the curvature $R$ of smectic shell and bend elastic constant $K_3$  play the role of the magnetic field or dilative strain in the Helfrich--Hurault instability in planar geometry~\cite{book:intro,napoli:2009}. By adopting a perturbation  approach with $h/R$ being a small parameter of our system, we explore the instability of the ground state (i) for $h\to0$ with respect to the periodic solution in thick smectic shells.  Our  prediction of the critical period and the critical thickness accounting for chevron-like texture can be relevant to analyze  experimental data and extract values of elastic constants and anchoring strength in the vicinity of the nematic--smectic phase transition. 

In the following we first reconsider the ground state of thin smectic shells accounting for splay-rich topological defects. Next,  we consider the breaking down of this solution state for thick smectic shells, assuming infinitely strong planar anchoring at the inner shell and neglecting the role of  splay rich topological defects ($K_1\ll K_3$).  In the last section we  hypothesize periodic texture of smectic layers and study the range of its stability.

%%%%%%%%%%%%%%%
\section{Ultrathin smectic shells}\label{sec:thin}
%%%%%%%%%%%%%%%
The orientational order of liquid crystals is described by the director $\bn$, which is a unit vector, confined to tangent plane of a sphere. The in-plane spatial variations of the director field  $\bn$ can be decomposed into splay $\nabla\cdot \bn$  and bend $\nabla \times\bn$ elastic distortions, where $\nabla$ is the covariant derivative. Thus the elastic  Frank free energy contains two independent contributions, integrated over surface of a sphere as~\cite{vitelli:2006,bowick:review,napoli:2012,frank:1958}
\be\label{eq:fel}
{\cal F}_{\rm el}=\frac 12\iint_{\cal S} dS\,\big\{K_1(\nabla\cdot \bn)^2+K_3|\nabla\times\bn|^2\big\}.
\ee
In case of nematic liquid crystals, the splay $K_1$ and bend $K_3$ elastic constants are of the same order $K_1\simeq K_3$. Therefore both splay and bend deformations are allowed and the ground state of ultrathin (2D) nematic shells contains four (+1/2) disclinations placed at the vertices of tetrahedron~\cite{vitelli:2006}. In the vicinity of nematic--smectic phase transition the elastic constant $K_3$ grows, and  bend  distortions are expelled from the sample, similar to the magnetic field in superconductors~\cite{cladis:1974}. The configuration of defects in smectic liquid crystal ($K_3\gg K_1$) confined to a sphere corresponds  to four (+1/2) disclinations lined up on a great circle of a sphere~\cite{bowick:prl}. Below we illustrate this point analytically.

\begin{figure}[tb]
\centering
\raisebox{43mm}{(a)}\includegraphics[height=45mm]{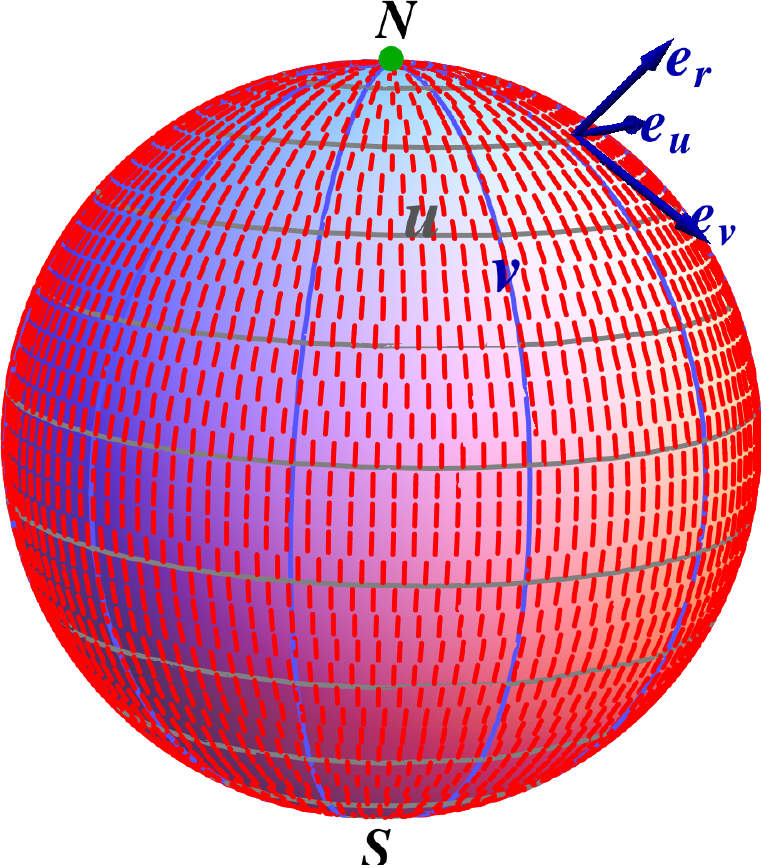}
\hskip2ex
\raisebox{43mm}{(b)\kern-10pt}\includegraphics[height=41mm,clip=true]{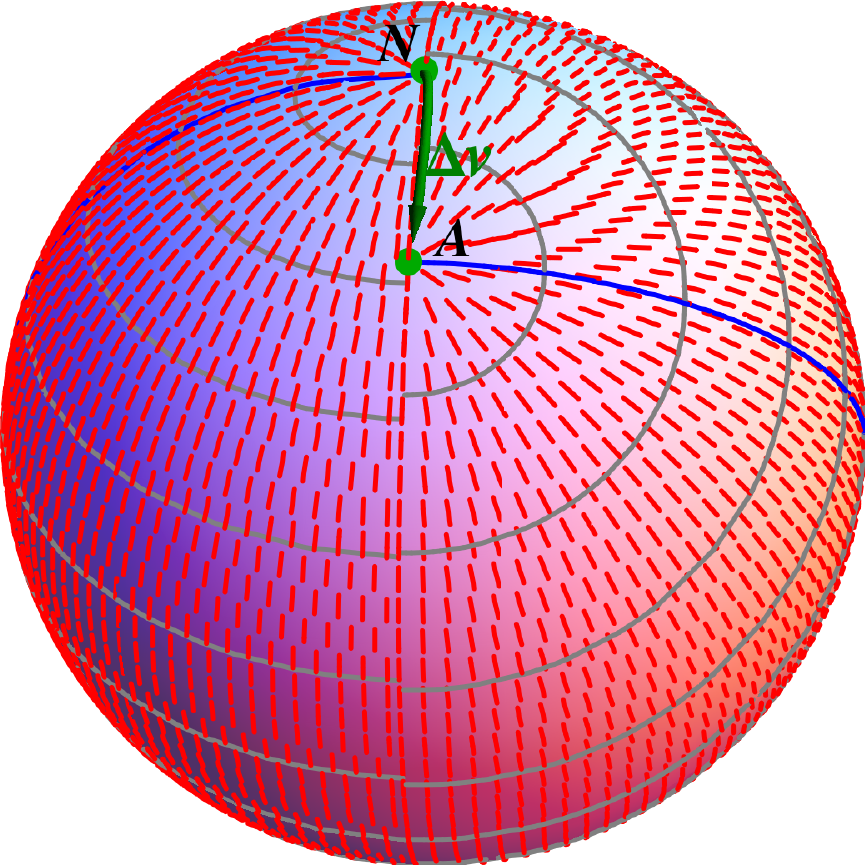}
\hskip1ex
\raisebox{43mm}{(c)\kern-10pt}\includegraphics[height=42mm]{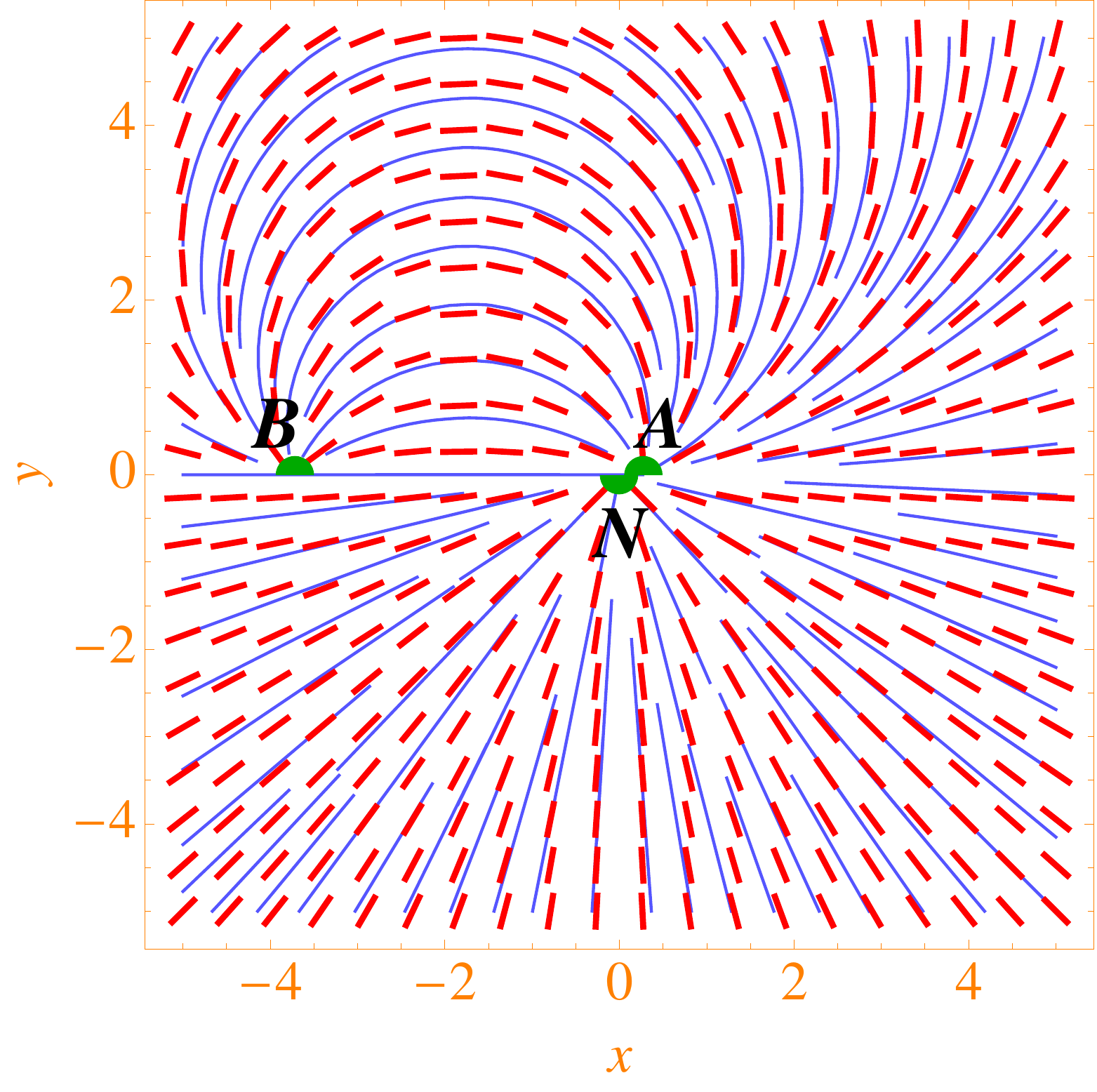}
\caption{(a) The director field $\bn$ (red) on a sphere aligned along meridians ($v$-lines) results into two $+1$ disclinations at the North ($N$) and South ($S$) poles.  Smectic layers are orthogonal to $\bn$ correspond to parallels ($u$-lines).  (b) Splitting of two defects by cut-and-rotate procedure of one hemisphere, does not cost any energy~\cite{bowick:prl} and results into four halves of $+1$ defect located at points $A,N$ and $B,S$ (other side of a sphere) separated by the angle $\Delta v=\pi/6$. (c) The director field in (b) projected from a sphere on a complex plane~\eqref{eq:zAB} with $z_A=\tan(\Delta v/2)$, $z_B=-\cot(\Delta v/2)$, $z_N=0$ and $|z_S|=+\infty$.}
\label{fig:sphere}
\end{figure}

Let us parametrize the director $\bn=\cos\alpha\,\bi_u+\sin\alpha\,\bi_v$, in local system of coordinates $(u,v)$,  with  sphere of radius $R$ being described by the radius vector $\br=R(\cos u\sin v, \sin u \sin v,\cos v)$. The simplest director configuration with $\alpha=\pi/2$  follows meridians (or $v$-lines)  connecting two $+1$ defects at the North  ($v=0$) and South ($v=\pi$) poles of a sphere as shown in Fig.~\ref{fig:sphere}a. Thus smectic layers described by the level set $\omega=R v=\const$  correspond to the parallels (or $u$-lines).  The director  $\bn=\nabla\omega/|\nabla\omega|=\bi_v$  is  locally orthogonal to smectic layers and can be defined everywhere, except at the two poles  ($v=0,\pi$), where it rotates by $+2\pi$. One can show explicitly that 
\begin{align}\label{eq:nabla1}
\nabla\cdot\bn&=\frac{\sin\alpha}{R\sin v}(\cos v-\p_u \alpha)+\frac{\cos\alpha}R\p_v\alpha,\\\label{eq:nabla2}
\nabla\times\bn&=-\frac{\cos\alpha}{R\sin v}(\cos v-\p_u \alpha)+\frac{\sin\alpha}R\p_v\alpha,
\end{align}
whence the configuration with angle $\alpha=\pi/2$ has  no bend ($\nabla\times\bn=0$). Note that, unlike the 3D case where the curl of a vector field is a vector (1-form), in the 2D  case the curl is defined as the Hodge dual of an exterior derivative applied to the 1-form, which gives a scalar (0-form)~\cite{book:flanders} or in index notation $\nabla\times\bn=\epsilon^{ij}D_i n_j$, where $\epsilon^{ij}$ is the dual of the Levi-Civita tensor~\cite{bowick:review}.  %%\marginpar{fine?}
Then the free energy~\eqref{eq:fel} has purely splay contribution 
\be\label{eq:falph}
{\cal F}_{\alpha=\pi/2}=\frac {K_1}2 \iint du\, dv\, \sin v\cot^2 v =2\pi K_1\bigg(\log\bigg(\frac{2R}a\bigg)-1\bigg),
\ee
where $a$ is related to the size of the defect core or  can be approximated by the size of 8CB molecule $\sim\!3$~nm. We expect that the energy contribution of the defect core is negligible here and thus can be ignored.

%%%%%%%%%%%%%   Figure 1

It was shown numerically in~\cite{bowick:prl}  and confirmed experimentally in~\cite{nieves:2012} that there is a set of degenerate ground states of a smectic liquid crystal on a sphere. Indeed,  in the limit $K_3\to \infty$ the bend free configuration is identical to the requirement of $\bn$ being aligned along geodesic lines  $(\bn\cdot\nabla)\bn=0$~\cite{bowick:prl,bowick:pnas,kamien:2009} or great circles on a sphere. The  geodesic on a sphere is given by parametric form $u=u_A +\arccos(\tan v_A \cot v)$, depicted as blue curve in Fig.~\ref{fig:sphere}. By cutting a sphere with two $+1$ defects (Fig.~\ref{fig:sphere}a with $\bn=\bi_v$)  and  rotating one hemisphere along a great circle by an arbitrary angle $\Delta v$ we obtain variety of states which can be transformed into one another with zero energy cost (see Fig.~\ref{fig:sphere}b).  The resulting director configuration on the Eastern hemisphere  can be  found by projecting two halves of $+1$ disclinations positioned at points $z_A=R \tan(\frac{\Delta v}2)$ and $z_B=-R\cot (\frac{\Delta v}2)$  (see Fig.~\ref{fig:sphere}c)
\be\label{eq:zAB}
\theta(z)=\Im\log\big[(z-z_A)(z-z_B)\big], \qquad z=x+iy
\ee
from the complex plane $z=x+i y= R \tan(\frac v 2 )e^{i u}$ onto a sphere so that $\alpha(u,v)=\pi/2-\theta(z)+\arg z$. Inserting this {\it ansatz} into~\eqref{eq:nabla1}, \eqref{eq:nabla2} we find  $\nabla\times \bn=0$ and the numerical value of the integrated splay contribution equals to~\eqref{eq:falph} whence independent of $\Delta v$. Thus, the ideal smectic configurations with equidistant layers on a sphere contain either two $+1$ defects at the poles (Figs.~\ref{fig:sphere}a) or 4 ($+1/2$) defects (Figs.~\ref{fig:sphere}b) on a great circle. This result holds as long as we neglect contributions coming from core size of defects and describe the system as two-dimensional shell with in-plane orientational order. Below we  study smectic  shells with finite thickness $h$ and analyze the deviation from the texture shown in Fig.~\ref{fig:sphere}a.

%%%%%%%%%%%%%%%%
\section{Finite thickness effects}\label{sec:thick}

First, we extend the bend free planar solution, found in the previous section  (see Fig.~\ref{fig:sphere}a), into the third direction  along the radius of a sphere $r\in[R,R+h]$, so that we get a smectic shell of finite thickness $h$. The non-vanishing curl of the director $\nabla\times\bi_v=\bi_u/r$ now results in a non-uniform layer spacing in the 3D smectic shell. In the vicinity of the nematic--smectic phase transition both twist $K_2$ and bend $K_3$ elastic constants, entering the conventional Frank free energy, diverge~\cite{cladis:1974}. For simplicity we assume that they are of the same order so   $K_2[\bn\cdot(\nabla\times\bn)]^2+K_3[(\bn\cdot\nabla)\bn]^2=K_3|\nabla\times\bn|^2$, thus the form of the elastic free energy is analogous to~\eqref{eq:fel}, except for the region of integration. Integrating over the finite thickness gives a bend contribution $2\pi h K_3$ and a splay contribution as~\eqref{eq:falph}, multiplied by the thickness $h$. Indeed, the dilation of smectic layers extendansatzd from a spherical surface along the normal is proportional to the increase of the area element  by $dS=dS|_{h=0}(1+ 2h H^2+h^2 K )$, where $H=1/R$ is the mean curvature  and $K=1/R^2$ is the Gaussian curvature  of an inner sphere.  To compensate the unfavored increase of interlayer spacing the director $\bn$ should tilt out of the tangent plane with non-zero gradient along the thickness, which follows from the expression for the curl of the vector field $\bn=n_r\bi_r+n_v\bi_v+n_u\bi_u$ written in spherical coordinates
\be 
\nabla\times\bn= \frac 1{r \sin v}\big[\p_v(n_u \sin v)-\p_u n_v\big]\bi_r+\frac 1{r\sin v}\big[\p_u n_r-\sin v \p_r(rn_u)\big]\bi_v+\frac 1r\big[\p_r(r n_v)-\p_v n_r\big]\bi_u.
\ee

\begin{figure}[t]
\centering
\raisebox{47mm}{(a)}\includegraphics[height=45mm]{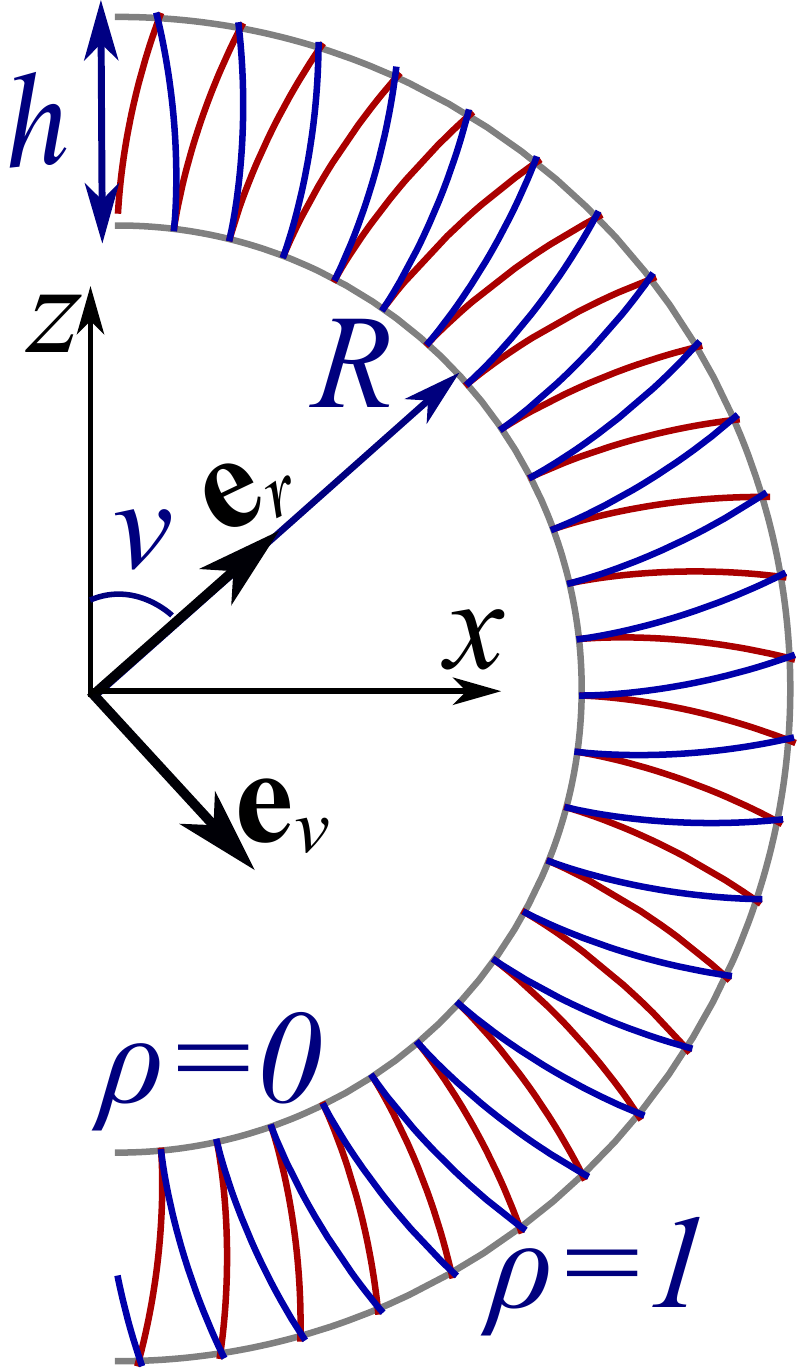}
\hfil
\raisebox{47mm}{(b)}\includegraphics[height=45mm]{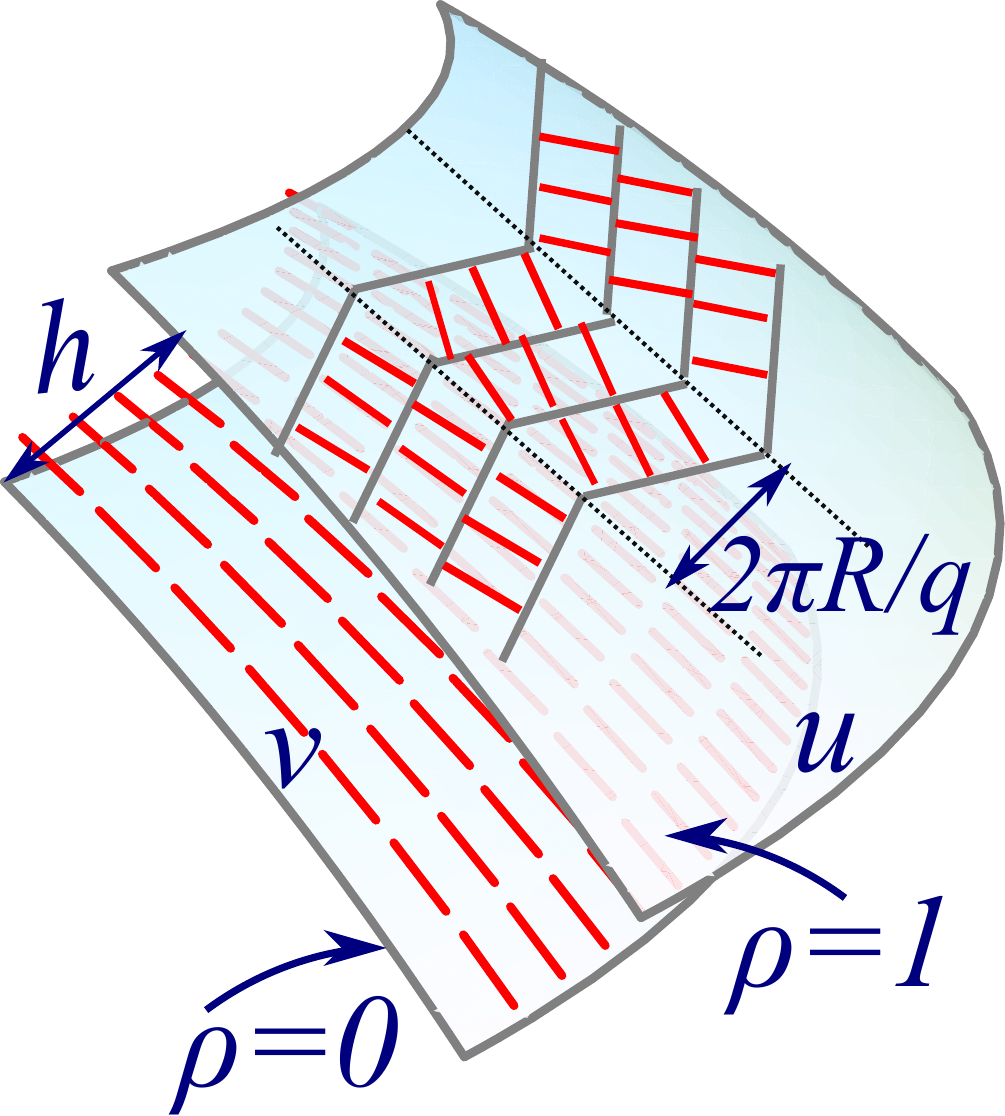}
\hfil
\raisebox{47mm}{(c)}\includegraphics[height=50mm]{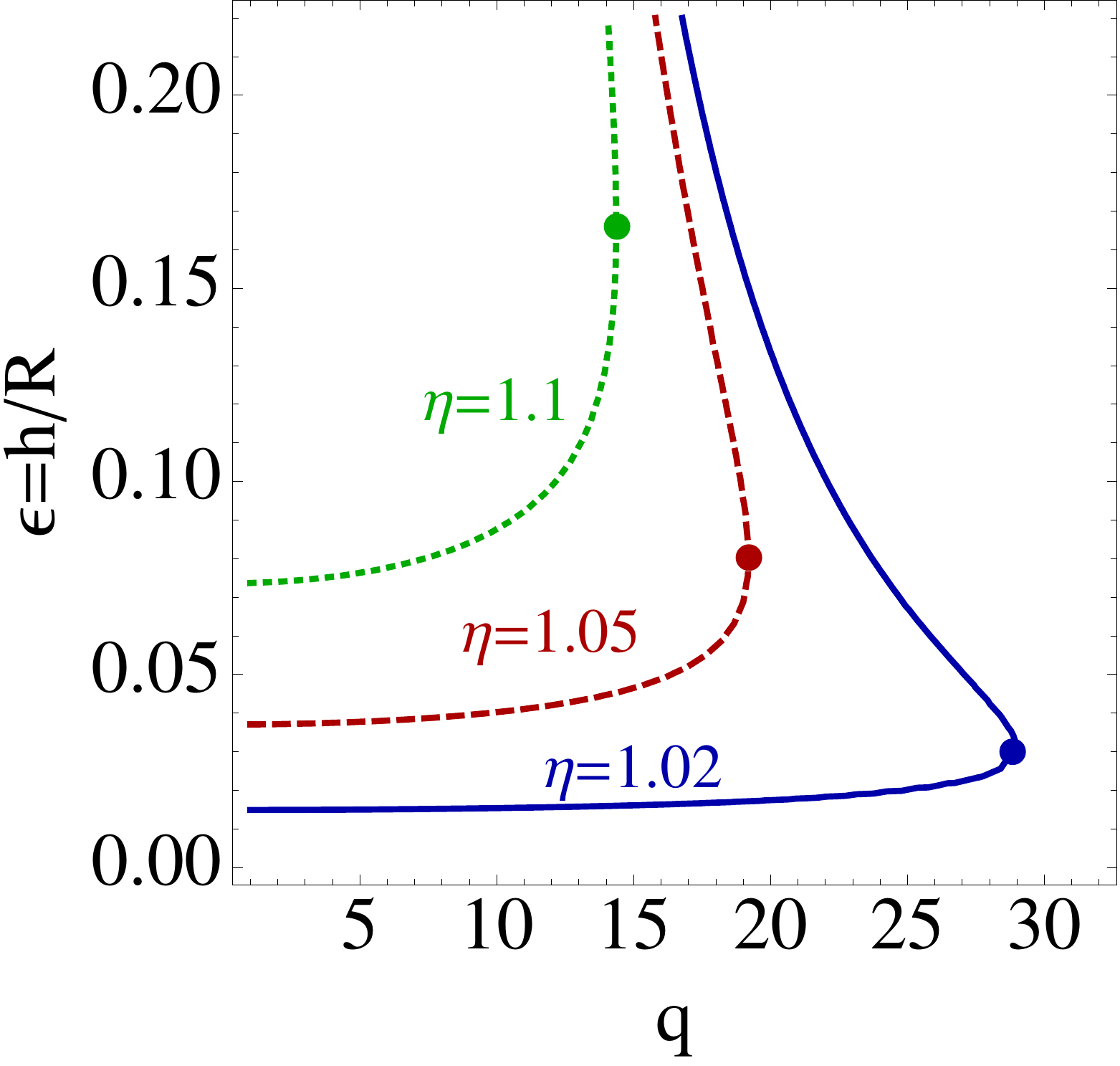}
\caption{Schematic illustration of the organization of smectic layers: (a) in $rv$-plane described by~\eqref{eq:rlay} with (counter)-clockwise tilt of the director corresponding to the (blue) red curves, $\eta=1.05$;   (b) within a segment of a spherical shell.  At the inner sphere $\rho=0$ the director (red) is aligned along meridians, while at the outer sphere $\rho=1$ the chevron texture has the highest tilt angle with non-zero projection on both $u$ and $r$ axes. (c) For $\eta=K_3/(RW_a)>1$ there is a critical  wavenumber $q_c$ (extremum of the curve), characterizing an instability from the state with uniform director $\bn$ and chevron structure with periodicity $2\pi/q$.   The tilt angle as well as the critical thickness grows with $\eta$. }
\label{fig:chevron}
\end{figure}

Let us introduce a small parameter $\ve=h/R\ll1$ and the rescaled variable $\rho\in[0,1]$, such as $r=R(1+\ve \rho)$. Then $\p_r=1/(\ve R)\p_\rho\sim O(\ve^{-1})$  is the leading order contribution to the gradient of a scalar function in spherical coordinates $\nabla\omega =\p_r\omega\, \bi_r+\frac {\p_v \omega}r \bi_v +\frac{\p_u\omega}{r\sin v} \bi_u$. Consider a layer perturbation in the form  $\omega=R v +\ve^2 R g(\rho)+O(\ve^4)$,  yielding
\begin{gather}\label{eq:bn1}
\bn=\frac{\nabla\omega}{|\nabla\omega|}=\ve \hat g\bigg(1+\rho\ve-\frac 12 \hat g\ve^2\bigg)\,\bi_r+\bigg(1-\frac 12 \hat g^2\ve^2-\rho \hat g^2\ve^3\bigg)\,\bi_v+O(\ve^4),\\
 %\frac{\p_\rho\omega}{\ve R|\nabla\omega|} \, \bi_r+\frac{\p_v\omega}{R(1+\ve\rho)|\nabla\omega|}\,\bi_v, \\
\nabla\times\bn=\bigg(\frac1 R-\frac{\rho+\hat g \hat g'} R \ve+\frac{2\rho^2-3 \hat g^2 -4\rho \hat g\hat g'}{2R}\ve^2\bigg)\,\bi_u +O(\ve^3),\label{eq:curln1}
 \end{gather}
where the normal to the layers coincides with the director $\bn$, which has two non-zero components and  accounts for the out-of-plane tilt $n_r\propto\hat g=\p_\rho g=g'$ or bend of the layers in $rv$-plane. We assume that bend deformations~\eqref{eq:curln1} are the dominant contribution to the free energy~\eqref{eq:fel} in the limit $K_3\gg K_1$ (more precisely $K_1/K_3\sim O(\ve^3)$). 
The presence of pure splay topological defects at two poles of a sphere as well as their structure in thick smectic shells is not accounted for here. To find $\hat g(\rho)$ minimizing the layer dilation we solve the Euler--Lagrange equation  within an {\it ansatz} $\hat g=\rho^m$
\be\label{eq:EL}
\hat g'^3+4\hat g \hat g ' \hat g'' +\hat g^2 \hat g'''=0, \quad \xrightarrow{m (6m^2 -7m+2)=0} \quad \hat g(\rho)=C_1 \rho^{1/2}+C_2 \rho^{2/3}.
\ee
According to experimental data~\cite{rudquist:2011} the inner shell $\rho=0$  imposes strong planar alignment and smectic layers are not distorted $\hat g|_{\rho=0}=0$, which is also suggested by~\cite{nieves:2012}.  At the outer shell ($\rho=1$) the undulations of the director are bound by the planar anchoring, written as  ${\cal F}_a= W_a/2 \int dS\,(\bi_r\cdot\bn)^2$, which penalizes the out-of-plane tilt of the director.   Up to the lowest order in $\ve$ the total free energy is $-2\pi R (C_1+C_2)^2\ve^2 (K_3-RW_a)$. The tilt $\hat g\neq 0$ is energetically favored if and only if $K_3>R W_a$. Otherwise, the in-plane alignment of the director $\bn=\bi_v$ ($\hat g=0$) uniform along the thickness is stable.  The next order approximation gives the critical thickness of smectic shell  $\ve_c\propto K_3/(R W_a)-1$ and the  amplitude of perturbation. Except for a trivial solution with $C_1=C_2=0$ we get  the physically irrelevant cases with complex amplitudes,  and two solutions with  $C_2=0$, given by
\be\label{eq:hatg}
\ve \hat g\simeq\pm \sqrt{2\ve \rho\bigg( 1 -\frac1 \eta \bigg)}, \qquad \eta=\frac{K_3}{R W_a}>1. 
\ee
Substituting this result in~\eqref{eq:curln1} we find up to the lowest order  
\be\label{eq:curln2}
(\nabla\times\bn)\cdot\bi_u=\frac 1{R \eta}+\frac{3-\eta-3\eta^2}{R \eta^2} \rho\ve+O(\ve^2).
\ee
As predicted above and follows from~\eqref{eq:curln2} the tilt of the director $\bn$~\eqref{eq:bn1}, \eqref{eq:hatg}, resulting in the bending of smectic layers in $rv$-plane illustrated in Fig.~\ref{fig:chevron}a, is energetically favoured for $\eta>1$. This transition from a uniform planar state appears in the vicinity of the nematic--smectic phase transition accompanied by the growth of $K_3$. For weak anchoring $W_a$ at the outer shell ($\eta\gg 1$) we get $\nabla\times\bn\sim -3\ve\rho/R\sim O(\ve)$, while for strong anchoring $\eta\geqslant 1$ the tilt angle~\eqref{eq:hatg} is suppressed. 
%Infinitely strong anchoring  $\eta\to0$ ($W_a\to \infty$) favours the base state, while for weak anchoring $\eta\to\infty$ ($W_a\to 0$)  we get~\eqref{eq:curln2}  $\nabla\times\bn\sim-3\rho\ve/R\sim O(\ve)$. Thus the  base state with the curl of $\bn$ of $O(1)$ becomes unstable and the dilation of smectic layers is indeed minimized by the out-of-plane tilt of the director when $\eta>1$ as predicted above. It follows from~\eqref{eq:hatg} that the absolute value of the tilt angle $\ve \hat g$ scales  with thickness as $\sqrt\ve$ and is suppressed by the finite anchoring $W_a$. Thus the tilt of the director $\bn$ sets in only above the threshold  $K_3>R W_a$, which occurs in the vicinity of nematic--smectic phase transition, consistent with experimental data~\cite{nieves:2012,nieves:2011,rudquist:2011}. 
Assuming  the anchoring strength $W_a\simeq 10^{-4}$~J/m$^2$ and the typical radius of a sphere $R\simeq 100~\mu$m~\cite{nieves:2012,nieves:2011,rudquist:2011}, one expects the critical value  $K_3 \gtrsim 10^{-8}$~J/m  for the instability threshold.  Above the threshold  $\eta=K_3/(R W_a)=1.05$ and for $\ve=h/R=0.04$ we get the estimate for the tilt angle as $\ve\hat g|_{\rho=1}\simeq \pm 0.062$.  In general, the solution~\eqref{eq:hatg} may be homogeneous with (``+'')``-'' sign, which is illustrated in Fig.~\ref{fig:chevron}a with smectic layers bending (counter)-clockwise in polar $rv$-plane, making use explicitly of our initial {\it ansatz}
\be\label{eq:rlay}
\omega= R v \pm \frac 2 3 R\sqrt{2(\ve\rho)^3\bigg(1-\frac 1\eta\bigg)}=\const \quad \xrightarrow{\ve\rho=r/R -1} \quad \frac r R=1 +\sqrt[3]{\frac{9 (\const-v)^2}{8(1-1/\eta)}},
\ee
Note that one could have chosen or guessed the scaling of $\omega$ with  the power  $\ve^{3/2}$,  but the results~\eqref{eq:EL}--\eqref{eq:rlay} do not change. One may guess that if the combination of both ``+'' and ``-'' signs in~\eqref{eq:rlay} is energetically allowed, then the change of sign of $\hat g$  likely results in the formation of curvature walls~\cite{blanc:1999}.  Because of the spherical geometry a na\"ive estimate of the number of curvature walls is $\pi/(\ve \hat g|_{\rho=1})$.  However, such a simple geometric argument based on the symmetry of the solution~\eqref{eq:hatg} may be irrelevant in accounting for the variation of birefringence  colors along $v$-lines connecting two defects in Fig.~\ref{fig:alberto}.  In fact, according to~\cite{nieves:2012,nieves:2011,rudquist:2011}   the modulation inside crescent domains is interpreted as a secondary instability.   In the next section we try to gain insight into the formation of the primary chevron-like periodic structure  along the $u$-direction.

%%%%%%%%%%%%%   Figure 2

%%%%%% 
\section{Pattern selection mechanism}\label{sec:pattern}

The main goal of this paper is not to explain the complexity of fully developed texture in smectic shells (e.g. Fig.~\ref{fig:alberto}) but to find the criterion, if there is one, for the onset of a periodic chevron-like solution. Again we adopt a simple perturbative approach and  allow both the tilt of the director $\bn$ in the tangent $(u,v)$-plane as well as variation along the thickness ($r$-direction). We assume the following form for a scalar function $\omega=R v +\ve^2 R g(\rho) u \sin v+O(\ve^4)$, which describes isosurfaces of smectic layers and accounts for chevron-like texture in $u$-direction with period $2\pi R/q$, where $q$ is unknown integer (see Fig.~\ref{fig:chevron}b). The base state with $g(\rho)=0$ is favoured by the planar anchoring, but becomes unstable  due to the growing bend contribution with thickness $h$, as was shown in section~\ref{sec:thick}.  The non-zero tilt $\nabla_u \omega\neq 0$ decreases the elastic energy contribution at the expense of the anchoring at the outer and possibly inner sphere. By optimizing the resulting free energy difference between the distorted state and base state, given by
\begin{multline}\label{eq:ftot2}
\frac{{\cal F}_{el}+{\cal F}_a}{4\pi^3}=\frac{K_3 R}{9q^2}\int_0^1 d\rho\,\bigg\{-{4 g' g''}\ve^2+
\frac{96 \pi^2}{25q^2} g'^2g''^2\ve^3-5 g'^2 \ve^3 -12 \rho  g'g''\ve^3 \bigg\}+\\+\frac{4W_ag'^2}{9q^2} \ve ^2\big (1 + {2  \rho }\ve\big)\Big|_{\rho=0,1}+O(\ve^4),
\end{multline}
we consider the interplay of bulk elastic term and boundary term, neglecting the role of splay contribution. The special solution of the Euler--Lagrange equation is  $g(\rho)=\pm {5 q \rho(\eta +\rho -1)}/{(8 \sqrt{6} \pi) }$, %$g(\rho) =C_1 \rho \pm 5q\rho^2/(8\sqrt 6 \pi)$, 
linear in the number of periods of the chevron pattern $q$ and quadratic in $\rho$, which differs from~\eqref{eq:hatg}.  The integration constant $C_1$ was determined from  the natural boundary condition at the outer shell $\p({\cal F}_{el}+{\cal F}_a)/\p C_1|_{\rho=1}=0$, while at the inner shell we assume  $g|_{\rho=0}=0$. As before we neglected the splay $K_1$-term in the free energy, which may be important to estimate the energy of curvature walls~\cite{blanc:1999}.

%Without experimental data about the director alignment at the inner sphere, we assume for simplicity that $g|_{\rho=0}=0$, while  the integration constant $C_1$ is determined from the natural boundary condition at the outer shell $\p({\cal F}_{el}+{\cal F}_a)/\p C_1|_{\rho=1}=0$. %The expression for $C_1$ is bulky to reproduce here.

In Fig.~\ref{fig:chevron}c the coexistence curve separates the region  where  the base state with unperturbed  director $\bn=\bi_v$  is stable from the region where the in- and out-of-plane tilt of the director with $g(\rho)\neq 0$ and $q>1$ within assumed {\it ansatz} is energetically preferred.  For $\eta=\const>1$ there is a critical point $(q_c,\ve_c)$, which identifies the wavelength $2\pi R/q_c$  and the thickness $h_c=\ve_c R$ at the instability threshold. We find that for  thin shells with $\ve_c\simeq 0.03$ the chevron texture is fine, corresponding to large $q_c$. For thicker shells the absolute value of the tilt angle $\ve_c g|_{\rho=1}$ as well as the period $2\pi R/q_c$ increases, so that the texture becomes more pronounced. Interestingly, assuming a different perturbation {\it ansatz} in terms of the spherical harmonics $\omega=R v+\ve^2 \rho^m \cos(q u)\sin^q v+O(\ve^4)$, we recover a similar prediction on the effective decrease of the equilibrium $q$ with the increase of the shell thickness ($\ve$), while the exponent $m\in(\frac 12,\frac 54)$ depends stronger on $\eta$. Our results suggest the validity of the perturbation approach in the vicinity of the nematic--smectic phase transition and are qualitatively consistent with experimental data~\cite{nieves:2012,nieves:2011,rudquist:2011} on the formation of crescent-like domains in thicker part of smectic shells.

%%%%%%%%
\section{Conclusions}
% what is done and where to go
%

In this paper we presented a phenomenological approach to treat smectic films confined between two spherical surfaces.  Within the continuum Frank free energy for liquid crystals and the assumptions made, we have shown that the ground state of two-dimensional films with director aligned along geodesics of a sphere, becomes unstable for thick films when $K_3>R W_a$. The instability is  driven by finite curvature of the sphere $1/R$ and the requirement of the bend-free  configuration in the vicinity of the nematic--smectic phase transition when $K_3$ diverges. The  resulting geometric frustration can be relieved by the tilt of the director out of the tangent plane of a sphere, as well as by in-plane undulations. The  constructed simplified solution for the plausible organization of smectic layers allows to identify the onset of the instability towards periodic chevron-like texture and provides the dependence of the critical wavelength on the  critical thickness $\ve=h/R$ of smectic shells. Future comparison between theory and experimental data~\cite{nieves:2012,nieves:2011,rudquist:2011} can be useful to extract the values of elastic constant $K_3$ and the anchoring strength~$W_a$ and eventually propose  the equilibrium 3D organizations of smectic layers within  thick spherical shells.

\section*{Acknowledgements}
This issue is dedicated to Professor Martine Ben Amar.  The present paper, in particular, would never have been possible without Martine who encouraged one of the authors (OVM) to study pattern formation in liquid crystals. The approaches in this paper are inspired by Martine's enthusiasm for exact solutions, complex analysis and perturbation theory. Her high scientific standards and the elegant way to tackle the most intricate and interesting scientific problems are highly appreciated.
The authors acknowledge financial support from the Soft Matter Program of Syracuse University.

%\section*{References}

\bibliographystyle{elsarticle-num} 
 \bibliography{ref_shells}

%\begin{thebibliography}{00}


%\bibitem{}

%\end{thebibliography}
\end{document}